\documentclass[%
 aip,
 jmp,%
 amsmath,amssymb,
 reprint,%
]{revtex4-1}

\usepackage{graphicx}
\usepackage{epstopdf}
\usepackage{dcolumn}
\usepackage{bm}

\begin{document}

\preprint{AIP/123-QED}

\title[Hard MHD limit in $1/3$ sawtooth like activity in LHD]{Hard MHD limit in $1/3$ sawtooth like activity in LHD}

\author{J. Varela}
 \email{jvrodrig@fis.uc3m.es}
\affiliation{Observatoire de Paris, 5 place Jules Janssen, F-92195 Meudon, France}
\author{K.Y. Watanabe}
\affiliation{National Institute for Fusion Science, Oroshi-cho 322-6, Toki 509-5292, Japan}
\author{S. Ohdachi}
\affiliation{National Institute for Fusion Science, Oroshi-cho 322-6, Toki 509-5292, Japan}
\author{Y. Narushima}
\affiliation{National Institute for Fusion Science, Oroshi-cho 322-6, Toki 509-5292, Japan}

\date{\today}

\begin{abstract}

The optimization of LHD discharges in inward-shifted configurations with $1/3$ sawtooth like activity is an open issue. These relaxation events limit the LHD performance driving a periodic plasma deconfinement. The aim of this study is to analyze the $1/3$ sawtooth like activity in plasmas with different stability properties to foreseen the best operation conditions and minimize its undesired effects. We summarize the results of several MHD simulations for plasmas with  Lundquist numbers between $10^5$ and $10^6$ in the slow reconnection regime, studying the equilibria properties during the onset of a chain of $1/3$ sawtooth like events. The research conclusions point out that the hard MHD limit can be reached in the inner plasma region after the onset of a strong $1/3$ resonant sawtooth like event and trigger a plasma collapse. The collapse can be avoided if the system remains in the soft MHD limit, namely in a regime with a pressure gradient and a magnetic turbulence below the critical values to drive the soft-hard MHD transition. In the soft MHD limit the system relaxations are the non resonant $1/3$ sawtooth like events or a weak version of the $1/3$ resonant sawtooth like events. A system relaxation in the soft MHD regime drives a minor plasma deconfinement that doesn't limit the LHD performance if the event periodicity is not very high.

\end{abstract}

\maketitle

\section{Introduction \label{sec:introduction}}

The LHD operation must avoid the onset of a system relaxations in the hard MHD regime because it can be the precursor of a plasma collapse \cite{1,2}. The collapse reduces the LHD performance driving a large energy leak and a partial deconfinement of the plasma \cite{3}. The optimization of the LHD discharges requires an operation regime without plasma collapses, namely the soft MHD limit. The operation regimen where a collapse can be driven it is called the hard MHD limit, therefore it is critical to avoid the transition between the soft and the hard MHD regimes \cite{4}. 

The transition from the soft to the hard MHD regimes is linked with the stability properties of the plasma and it is triggered if the pressure gradient and the magnetic turbulence reach a critical value \cite{5,6}. The width of the magnetic islands is large in the hard MHD limit and an instability can drive a strong overlapping between islands. If a wide stochastic region covers the plasma a collapse can be driven \cite{7,8,9}. If the stochastic region overlies only a portion of the plasma, a minor collapse is driven.

The LHD inward-shifted configurations \cite{10}, magnetic axis in the vacuum shifted to the interior of the torus, are unstable to resistive MHD pressure-gradient-driven modes \cite{11,12}. There is a magnetic hill near the magnetic axis and the unstable modes are not stabilized by the magnetic shear \cite{13}, as it was reported in previous MHD linear studies \cite{14,15}. These instabilities increase the energy leaks of the system but they don't limit the LHD performance \cite{16}. These plasma relaxations are driven in the soft MHD limit because there is a stabilizing mechanism that avoid a strong destabilization of the low $n$ modes \cite{17,18}, but its efficiency decreases in operations with $ \beta_{0} > 1 \% $ \cite{19}. This mechanism consists in the evolution of the pressure to a staircase-like profile. The mode growth rate saturates in the profile flattening showing periodic excitations and relaxations. In the case of operations with high $ \beta_{0}$ values, the system transits to the hard MHD limit if there is a strong interaction between modes of different helicities and a plasma collapse can be triggered \cite{20}.

We analyze a LHD discharge in the inward-shifted configurations for a plasma fuelled by pellets and heated by strong NBI injection \cite{21}. The pressure profile is peaked and there isn't a large net toroidal current \cite{22}. Sawtooth like events are driven periodically by the unstable modes $n/m = 1/2$ and $1/3$ \cite{23,24}.

Previous investigations pointed out that this LHD configuration remains in the soft MHD limit for the instabilities driven in the middle plasma region \cite{25}. Only the $1/2$ sawtooth like events are driven, not its version in the hard MHD limit called internal disruption \cite{26,27}. The plasma conductivity and density are large enough to avoid the transition to the hard MHD regime \cite{28,29,30}. In the inner plasma region, two different $1/3$ sawtooth like events were analyzed, the non resonant and the resonant versions \cite{4}. The non resonant event is driven in the soft MHD limit and its effect in the device performance is small, a minor energy leak and a weak distortion of the plasma equilibria. The resonant event is a stronger relaxation and can be driven in the hard MHD limit.

The aim of this study is to analyze the equilibria properties during the transition between the soft-hard regimes in the inner plasma, as well as the role of the resonant $1/3$ sawtooth like events if a collapse is triggeed. The study conclusions are useful to foreseen the best operation conditions, avoiding or reducing the undesirable effects of the sawttoth like activity in the LHD performance. This is an important task for present and future LHD configurations because the $1/3$ sawtooth like activity is an open issue in the optimization of the operation model.

We simulate a chain of non resonant and resonant $1/3$ sawtooth like events for plasmas with different stability properties. Each simulation has a different value of the Lundquist number ($S$) and the operation regime can evolve from the soft to the hard MHD limit.

The simulations are made using the FAR3D code \cite{31,32,33}. We use a set of reduced non-linear resistive MHD equations to study the evolution of a perturbed VMEC equilibria \cite{34}. The equilibria is obtained before the onset of a sawtooth like event and after the last pellet injection. The equilibria is reconstructed using the electron density and temperature profiles calculated from the Thomson scattering and electron cyclotron emission data.

This paper is organized as follows; numerical model and equilibria characteristics in section \ref{sec:model}, simulation results in section \ref{sec:simulation} and  conclusions in section \ref{sec:conclusions}.

\section{Numerical model and equilibria characteristics \label{sec:model}}

The reduced set of MHD non-linear resistive equations without averaging in the toroidal angle is deduced for configurations with high-aspect ratio, moderate $\beta$ values (of the order of the inverse aspect ratio $\varepsilon=a/R_0$), small variation of the fields and small resistivity \cite{35}. This formulation solves the equations in an exact three dimensional equilibrium including the effect of the linear helical mode coupling. 

The code variables have two components, the equilibrium and the perturbation terms, $ A = A_{eq} + \tilde{A} $ with $ A_{eq} > \tilde{A} $. The perturbation term of the velocity and the magnetic field are

\begin{equation}
 \mathbf{v} = \sqrt{g} R_0 \nabla \zeta \times \nabla \Phi, \quad\quad\quad  \mathbf{B} = R_0 \nabla \zeta \times \nabla \psi,
\end{equation}
where $\zeta$ is the toroidal angle, $\Phi$ is a stream function proportional to the electrostatic potential, and $\psi$ is the perturbation of the poloidal flux.

The equations, in dimensionless form, are
\begin{equation}
\frac{{\partial \psi }}{{\partial t}} = \nabla _\parallel  \Phi  + \frac{\eta}{S} J_\zeta
\end{equation}
\begin{eqnarray} 
\frac{{\partial U}}{{\partial t}} = - {\mathbf{v}} \cdot \nabla U + \frac{{\beta _0 }}{{2\varepsilon ^2 }}\left( {\frac{1}{\rho }\frac{{\partial \sqrt g }}{{\partial \theta }}\frac{{\partial p}}{{\partial \rho }} - \frac{{\partial \sqrt g }}{{\partial \rho }}\frac{1}{\rho }\frac{{\partial p}}{{\partial \theta }}} \right) \nonumber\\
 + \nabla _\parallel  J^\zeta  + \mu \nabla _ \bot ^2U
\end{eqnarray} 
\begin{equation}
\label{peq}
\frac{{\partial p}}{{\partial t}} =  - {\mathbf{v}} \cdot \nabla p + D \nabla _ \bot ^2p + Q
\end{equation}
The vorticity is defined like $U =  \sqrt g \left[{ \nabla  \times \left( {\rho _m \sqrt g {\bf{v}}} \right) }\right]^\zeta$, where $\rho_m$ is the mass density. The lengths are normalized to the generalized minor radius $a$ and the time to the poloidal Alfv\' en time $\tau_{hp} = R_0 (\mu_0 \rho_m)^{1/2} / B_0$. The resistivity $\eta$, magnetic field $B$ and pressure $p$ are normalized to their averaged value in the magnetic axis. The Lundquist number $S$ is the ratio of the resistive time $\tau_R = a^2 \mu_0 / \eta_0$ to the poloidal Alfv\' en time.

The perpendicular dissipation terms are the collisional cross-field transport coefficient $D$ and the collisional viscosity coefficient for the perpendicular flow $\mu$. The system energy losses by numerical dissipation are balanced with the factor $Q$, added to the equation (\ref{peq}).

The system geometry is defined by the equilibrium flux coordinates $(\rho, \theta, \zeta)$. The Boozer coordinates simplifies the analysis and the Jacobian of the transformation is $\sqrt g$ \cite{36}. The generalized radial coordinate $\rho$ is proportional to the square root of the toroidal flux function and it is normalized to one at the edge. 

The operator $ \nabla_{||} $ is the derivation in the direction parallel to the magnetic field, defined like

\begin{equation*}
\nabla_{||} = \frac{\partial}{\partial\zeta} + \rlap{-} \iota\frac{\partial}{\partial\theta} - \frac{1}{\rho}\frac{\partial\tilde{\psi}}{\partial\theta}\frac{\partial}{\partial\rho} + \frac{\partial\tilde{\psi}}{\partial\rho}\frac{1}{\rho}\frac{\partial}{\partial\theta},
\end{equation*}
where $\rlap{-} \iota$ is the rotational transform.

The radial dimension is discretized using finite differences and the angular dimensions by Fourier expansions. The numerical scheme is semi-implicit in the linear terms and the non linear term are added explicitly. The nonlinear version uses a two semi-steps method to ensure $(\Delta t)^2$ accuracy.

\subsection{Equilibrium properties}

\begin{figure}[h]
\centering
\includegraphics[width=0.35\textwidth]{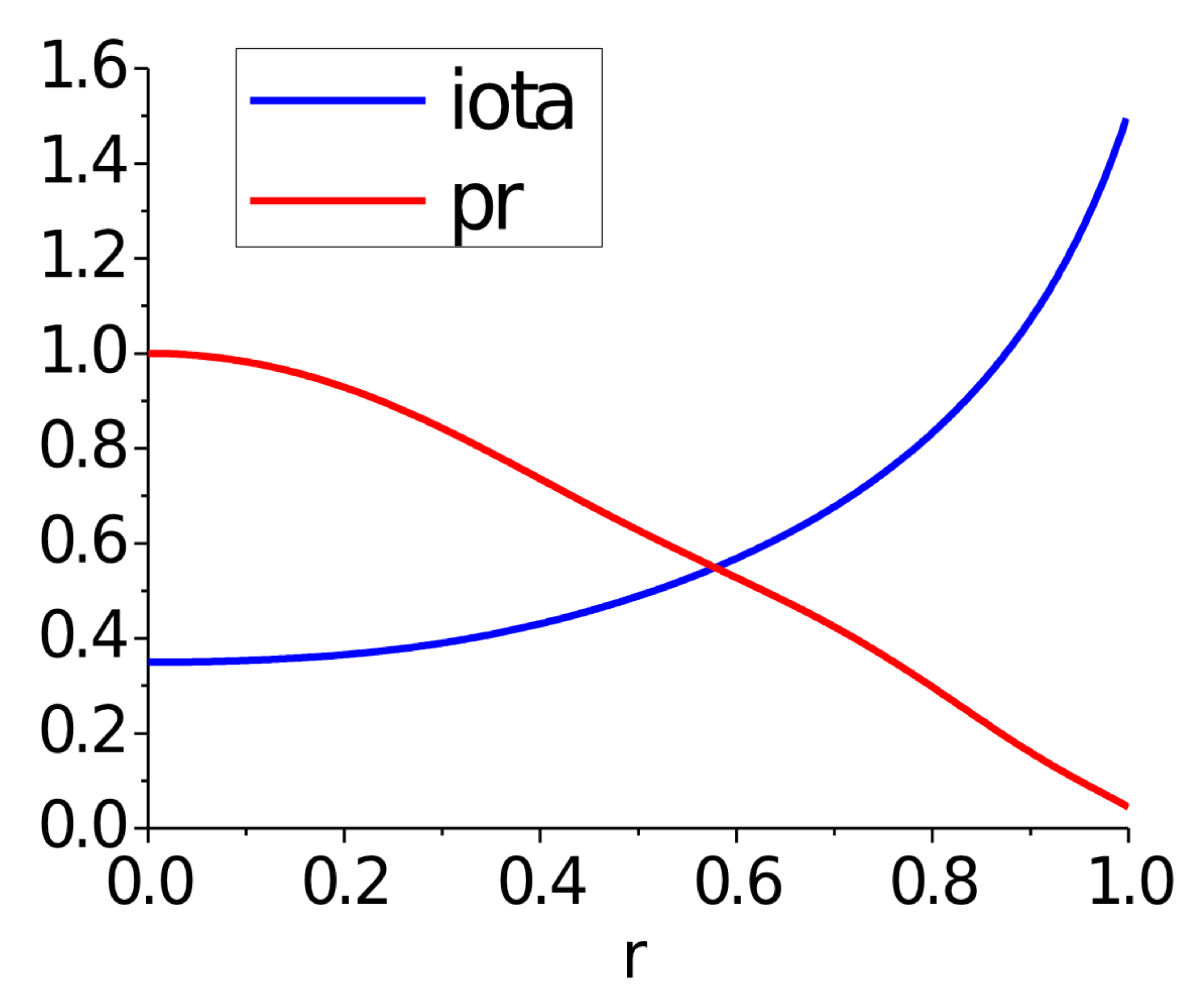}
\caption{Equilibrium pressure and rotational transform profiles.} \label{FIG:1}
\end{figure}

The plasma equilibria is calculated with a free-boundary version of the VMEC code. It is a high density plasma produced by sequentially injected hydrogen pellets and strongly heated by 3 NBI after the last pellet injection. The vacuum magnetic axis is inward-shifted ($R_{{\rm{axis}}} = 3.6$ m), the magnetic field at the magnetic axis is $2.75$ T, the inverse aspect ratio $\varepsilon$ is $0.16$, and the $\beta_0$ is $1.48 \%$. The figure~\ref{FIG:1} shows the equilibrium pressure and rational transform profiles.

\subsection{Simulation parameters}

The radial grid is uniform and it has 500 points. The toroidal $n$ and poloidal $m$ numbers are chosen to include in the simulation the most energetic $n/m$ modes and their most important toroidal coupled modes. For the equilibrium modes, the $n = 0$ family with the poloidal numbers $0 \le m \le 5$, is enough to reproduce the LHD shaping in first approximation. The selection of the dynamic modes is different in the simulation with low Lundquist number ($S=10^5$) than in the simulations with higher Lundquist number ($S=5 \times 10^5$ and $10^6$), because the simulations with higher $S$ numbers require an improved mode resolution in the plasma periphery to avoid numerical overshots and improve the simulations convergence. The low $S$ simulation has 511 dynamic modes for the $n = 1$ to $30$ toroidal families, but in the high $S$ simulations there are 493 dynamic modes for the $n = 1$ to $20$ toroidal families. The number of toroidal families in the high $S$ simulations is smaller to reduce the computational time.

The Lundquist number is around two orders lower than the experimental value for computational reasons. The consequence is a plasma more resistive and the simulation events will be stronger than the activity of the experiment, but the driver is in both cases a MHD resistive mode \cite{37}. We assume that the simulation is in the slow reconnection regime defined by the Sweet-Parker theory. The system remains in the slow reconnection regime if the Lundquist number is below a critical value. In this regime the current sheet is stable or marginal unstable to the super-Alfv\'enic plasmoid instability and the fast reconnections are not driven \cite{38,39,40,41}. The single fluid description of the present study is not valid in the fast reconnection regime. In our case the reconnection can be driven only via the Ohm's term, but in collisionless plasmas it is negligible compared with the electron inertia, Hall effect and the electron viscosity terms. To reproduce the Physics in the fast reconnection regime the simulation should be done at least with a two fluid model. The Lundquist number of a LHD plasma in the inner plasma is between $S = 10^7$ and $10^8$, a plasma in the fast reconnection regime, therefore this study cant reproduce the Physics of the reconnection regime, but we consider than the effect of the $1/3$ mode in the transition between the soft-hard MHD regimes is the same regardless of the reconnection rate.

The coefficients of the dissipative terms are $\mu=7.5 \times 10^{-6}$ and $D=1.25 \times 10^{-5}$. They are normalized to $a^2/\tau_{hp}$.

The source term $Q$ (\ref{peq}) simulates the strong NBI heating like a Gaussian centered near the magnetic axis, at $\rho = 0.2$, with a standard deviation of $\sigma = 0.15$. The energy input is dynamically fitted to keep constant the volume integral of the pressure.

We define a magnetic island between the grid point 175 and 288 like the initial condition of the perturbation in the VMEC equilibrium, therefore the system evolution begins from a nonlinear state.

\section{Simulation results \label{sec:simulation}}

This is a summary of three simulation with $1/3$ sawtooth like activity for plasmas with different Lundquist numbers, $S = 10^5$, $5 \times 10^5$ and $10^6$. We choose the strongest and more didactic chains of non resonant and resonant sawtooth like events. The stability properties of each plasma are different because the resistive modes in a simulation with a high $S$ value is less unstable than in a low $S$ case, therefore the simulation with lower $S$ value can transit from the soft to the hard MHD regime easily. 

The properties of the plasma equilibria during the soft-hard MHD transition is analyzed using the next diagnostics:

1) The magnetic energy (ME) of the system; it shows if an instability is under develop and when it reaches its maximum activity (local maximum of the profile). If the profile slope is sharp and the local maximum high, the instability is strong. 

2) The system energy loss; a drop of the profile points out that the system is losing energy. If the instability is strong the energy decay is fast and large.

3) The magnetic energy of the dominant modes; a correlation between the energy of the modes shows the plasma region destabilized by the perturbation. If the magnetic energy of several modes has a local maximum at the same time, the instability affects a wide plasma region. This diagnostic is useful too to know the mode that triggers the instability. 

4) The Magnetic turbulence; it is proportional to the perturbed magnetic field $\left | \widetilde{B} \right | / B_{0}$ . The width of the magnetic islands is large if the magnetic turbulence is high, therefore a plasma with large magnetic turbulence can transit easily to the hard MHD regime.

5) Pressure gradient; it is calculated in different location along the normalized minor radius, at $\rho = 0.1$, $0.3$, $0.5$ and $0.7$. A strong oscillation and a high local maximum of the pressure gradient show the plasma region where the perturbation is driven and the strength of its destabilizing effect in the plasma equilibria.

6) The averaged pressure; it is defined as $\left\langle p \right\rangle = p_{\rm{eq}}(\rho) + \tilde{p}_{00} (\rho)$ (the angular brackets indicate average over a flux surface and $\tilde{p}_{00}$ is the $(n=0, m=0)$ Fourier component of the pressure perturbation). The pressure profile is flattened near the rational surface of an unstable mode. If the profile has several small flattening, the destabilizing effect of the unstable modes is local, but if the profile shows only a few flattening driven by several unstable modes at the same time, the instability effect is non local. A collapse can be driven when a profile flattening covers a wide region the plasma.

7) The instantaneous rotational transform; it indicates the instantaneous position of the rational surfaces. A large deformation of the profile shows a plasma region strongly destabilized. It is defined like

\begin{equation}
\label{iota}
\rlap{-} \iota (\rho)+ \tilde{\rlap{-} \iota}(\rho) = \rlap{-} \iota+ \frac{1}{\rho}\frac{\partial\tilde{\psi}}{\partial\rho}
\end{equation}

8) Two-dimensional contour plots of the pressure profile; it is useful to trace the plasma regions with large gradients. There are two versions, the plots with the perturbed pressure $\tilde{p} = \sum_{n\not=0,m} \tilde{p}_{n,m}(\rho)cos(m\theta + n\zeta)$ and the plots with the full pressure $p = p_{eq}(\rho) + \sum_{n,m} \tilde{p}_{n,m}(\rho)cos(m\theta + n\zeta)$. The plot of the perturbed pressure show the evolution of the instability and the full pressure plot the structure of the flux surfaces. 

9) The Poincar\'e plots of the magnetic field structure; it is helpful to visualize the instantaneous topology of the magnetic field. It is calculated following the magnetic field lines around the torus. There are two different plots, one with only the dominant modes and another with all the modes of the simulation. The first type give information of the distribution and shape of the magnetic islands and the other about the stochastic regions in the plasma \cite{7}. If the width of the magnetic islands is large and they are strongly overlapped, large stochastic regions appears in the plasma. This method overestimates the size of the stochastic regions, the important information in these plots is the evolution of the stochastic regions during the events.

10) Plasma emissivity; it is proportional to the squared value of the pressure along a measurement chord, expressed like $ I = \int dl p^2 $ where $ dl = \sqrt{dR^2 + dZ^2} $ with $R$ the mayor radius and $Z$ is the height in real LHD coordinates. To calculate the emissivity we used the full pressure value, defined at 8), and the coordinates $(R,Z)$ before the Boozer transformation. We calculated the emissivity over the same measurement chords than a soft X ray camera in LHD to do a qualitative comparison between the plasma emissivity of the simulation and the experimental data. The real plasma has a poloidal rotation but in the simulation there aren't the terms to drive it, like the $\vec{E} \wedge \vec{B}$ or the diamagnetic current, therefore we add and artificial solid rigid rotation. The rotation is $v_{\theta} = 0.75$ km/s in the electron diamagnetic direction as the theoretical models suggest \cite{42}\cite{43}. 

\subsection{Simulation with $S = 10^5$}

We study a chain of a non resonant (I, blue line) and a resonant (II, red line) $1/3$ sawtooth like event. The Fig. 2 shows the magnetic energy evolution for the dominant modes. During the non resonant event the ME of the mode $1/3$ and $2/5$ are still increasing but in the resonant event they drop after reaching a local maximum. The ME of the $1/2$ mode remains almost constant during both events pointing out that the instability affects mainly the inner plasma region ($\rho < 0.5$).

\begin{figure}[h]
\centering
\includegraphics[width=0.48\textwidth]{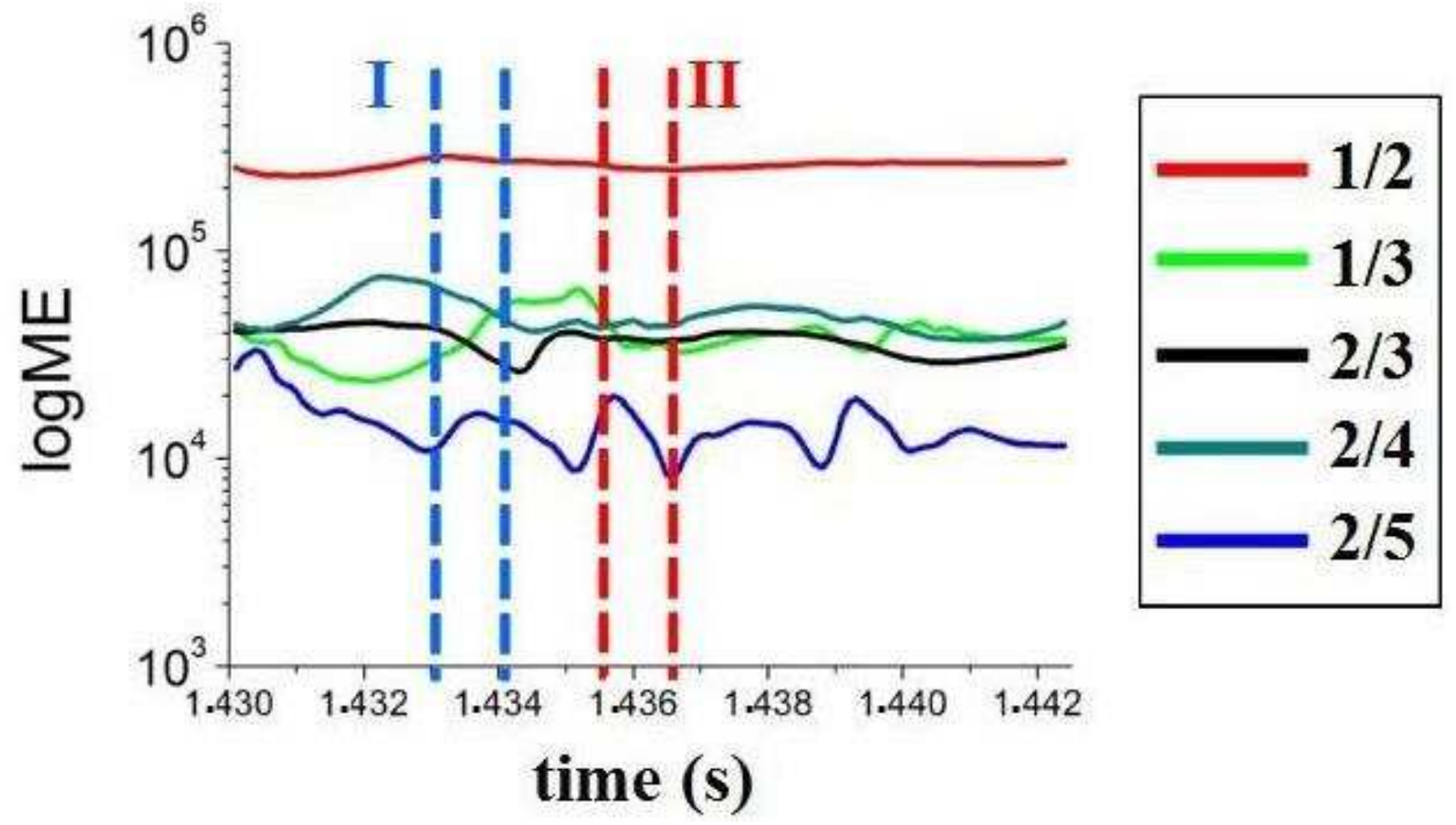}
\caption{Magnetic energy evolution of the dominant modes in the simulation with $S = 10^5$. The blue line shows the non resonant event and the red line the resonant event.} \label{FIG:2}
\end{figure}

In the non resonant event the dominant flattening of the pressure profile is in the middle plasma and it is driven by the $1/2$ mode (Fig. 3A). During the resonant event there are another two profile flattening in the inner plasma driven by the modes $2/5$ and $3/8$ around $\rho = 0,3$, and by the mode $1/3$ near the magnetic axis (Fig. 3C). The iota profile is deformed near the magnetic axis during both events and it drops below $\rlap{-} \iota = 1/3$ in the resonant case (Fig. 3D). The $1/3$ mode enters in the plasma during the resonant event, at $t = 1.4360$ s, very close to the magnetic axis, at $\rho = 0.04$, but remains outside the plasma in the non resonant case.

\begin{figure}[h]
\centering
\includegraphics[width=0.47\textwidth]{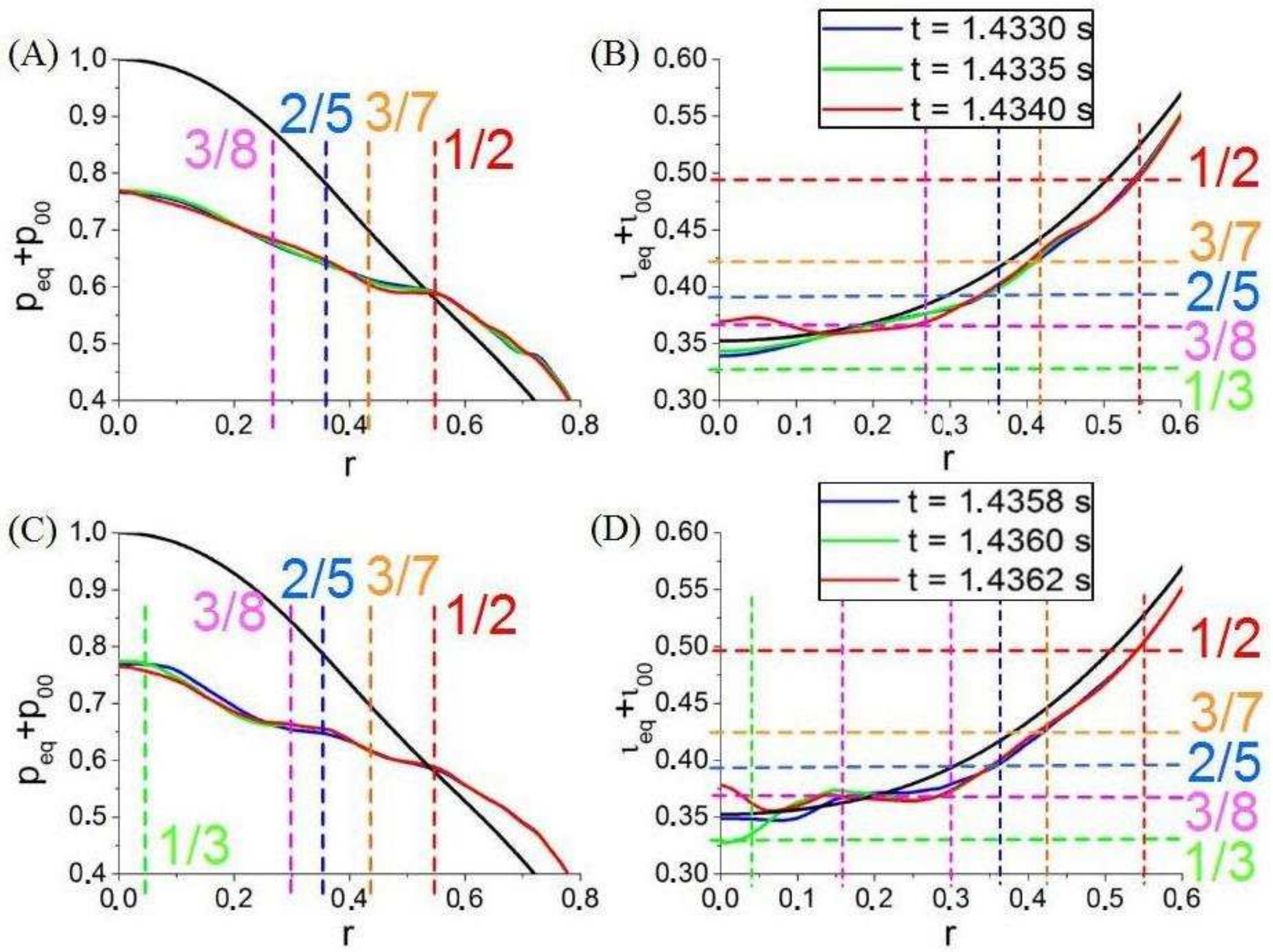}
\caption{Pressure (A and C) and rotational transform (B and D) profiles of the events I and II in the simulation with $S = 10^5$.} \label{FIG:3}
\end{figure} 

The contour plots of the perturbed pressure for the non resonant event show a perturbation with origin in the middle plasma which grows until it reaches the inner plasma region (Fig. 4A). In the resonant case there are two instabilities, one in the middle plasma and another near the magnetic axis. The perturbation in the inner plasma enhances but it is not linked with the perturbation in the middle plasma (Fig. 4B).

\begin{figure}[h]
\centering
\includegraphics[width=0.45\textwidth]{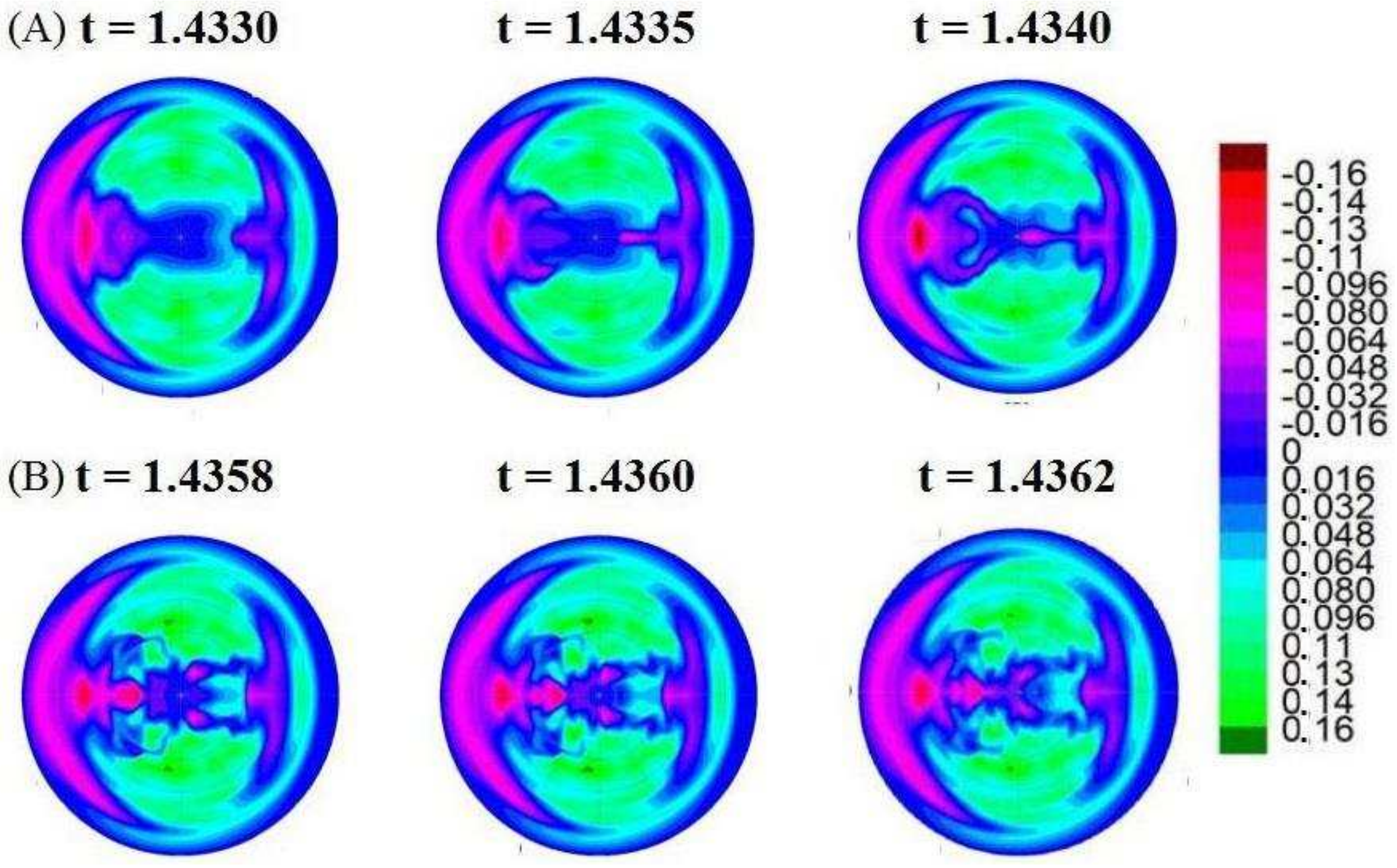}
\caption{Contour plot of the perturbed pressure in the simulation with $S = 10^5$.} \label{FIG:4}
\end{figure} 

During the non resonant event the magnetic surfaces are distorted near the magnetic axis and the width of the magnetic islands in the inner plasma slightly increases (Fig. 5A). The stochastic region in the middle plasma grows but it doesn't reach the nearby of the magnetic axis (Fig. 5B). In the case of the resonant event the magnetic surfaces in the inner plasma are torn and there are three magnetic islands close to the magnetic axis (Fig. 6A). There is a stochastic region covering the inner plasma that almost reaches the magnetic axis, but it is not linked with the stochastic region in the middle plasma (Fig. 6B).

\begin{figure}[h]
\centering
\includegraphics[width=0.4\textwidth]{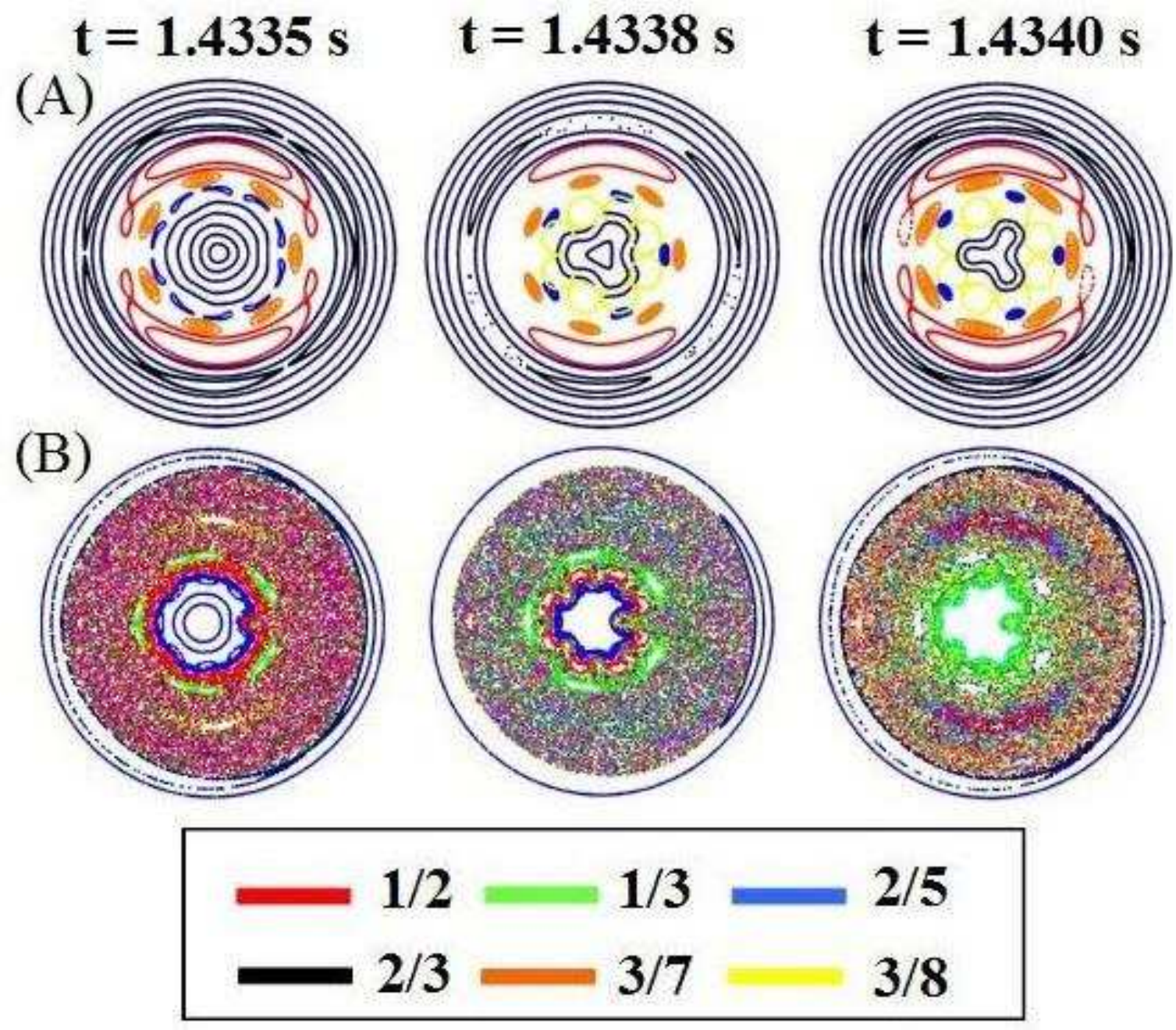}
\caption{Magnetic islands (A) and stochastic region (B) during the non resonant event in the simulation with $S = 10^5$.} \label{FIG:5}
\end{figure} 

\begin{figure}[h]
\centering
\includegraphics[width=0.4\textwidth]{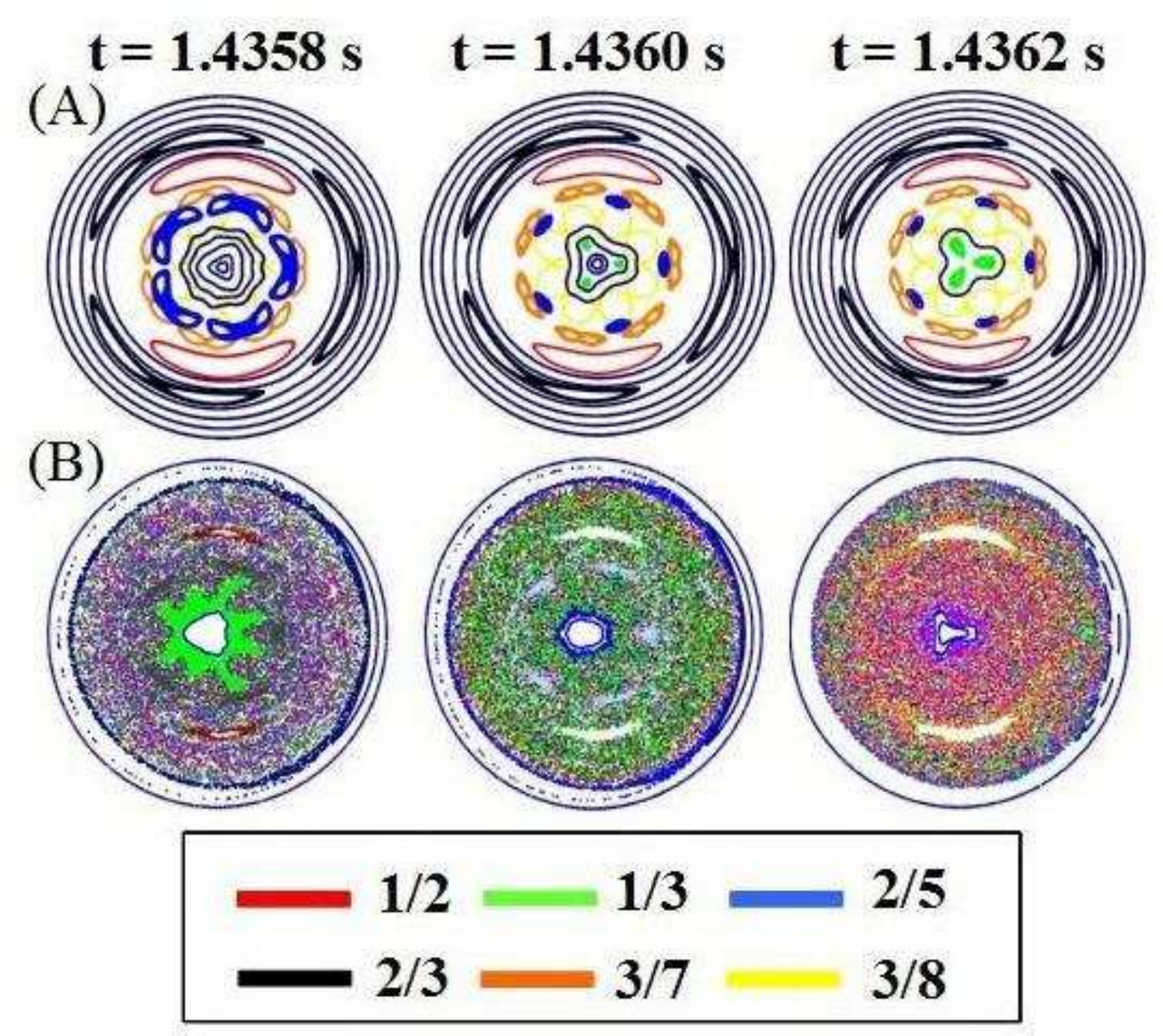}
\caption{Magnetic islands (A) and stochastic region (B) during the resonant event in the simulation with $S = 10^5$.} \label{FIG:6}
\end{figure} 

The emissivity in the non resonant event drops between the chords $\rho = 0.2$ and $0.3$ while the perturbation is growing, but it begins to increase as soon as the instability reaches the inner plasma region driving a drop of the emissivity in the chord $\rho = 0.1$ (Fig. 7). The emissivity in the chord $\rho = 0.4$ shows a small positive peak and the chords at $\rho > 0.4$ are not affected. This point out that the non resonant event is a local plasma relaxation. For the resonant event there is a drop of the emissivity in all the chords of the inner plasma at the same time, therefore it is a global relaxation that affects all the inner plasma region.

\begin{figure}[h]
\centering
\includegraphics[width=0.35\textwidth]{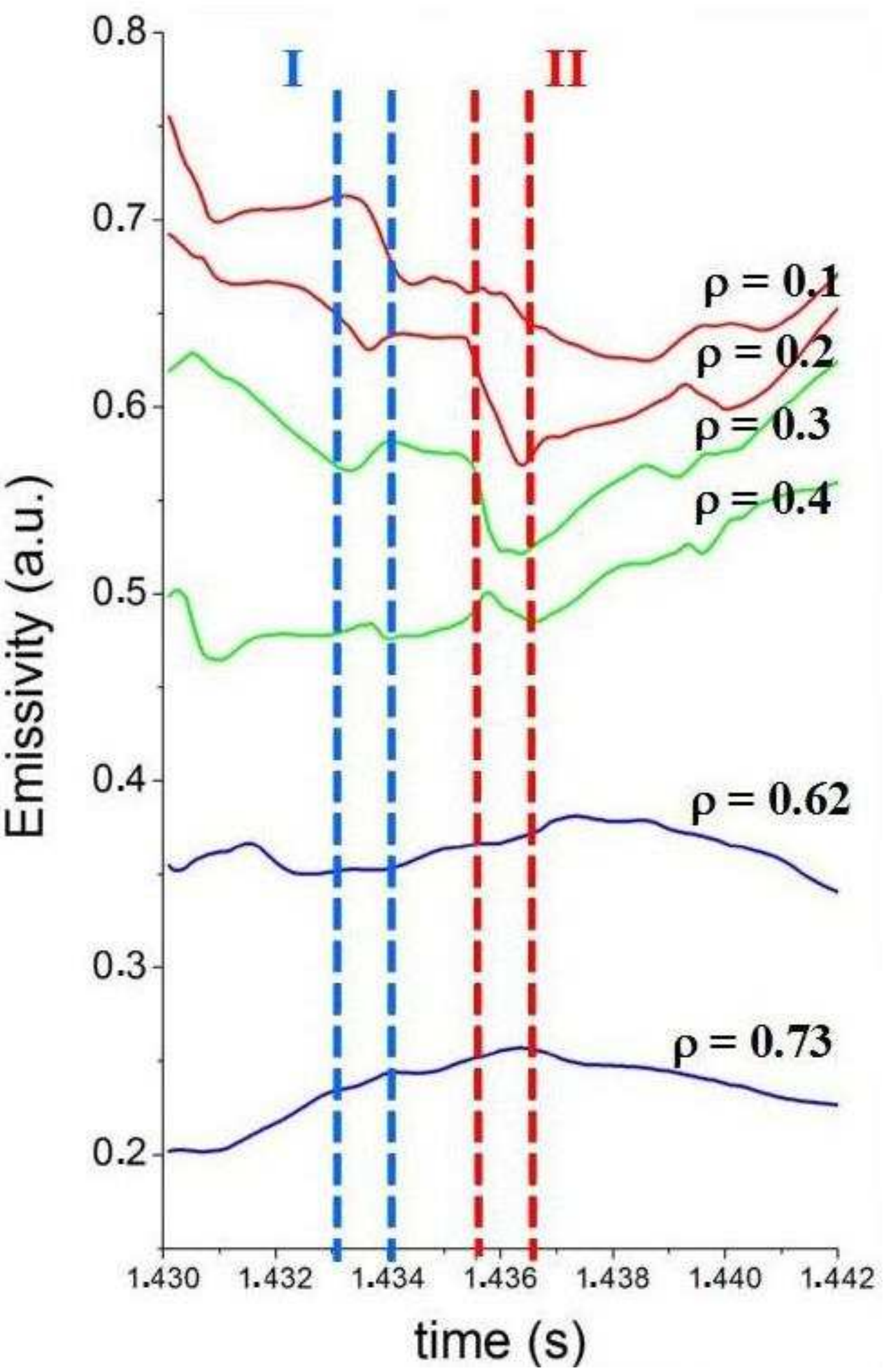}
\caption{Plasma emissivity in the simulation with $S = 10^5$.} \label{FIG:7}
\end{figure} 

These results show that the resonant event is driven in the inner plasma, close to the magnetic axis. The instability affects all the inner plasma region so it is a global relaxation. The magnetic islands of the modes $1/3$, $3/8$, $2/5$ and $3/7$ are overlapped and the stochastic region covers a wide region of the inner plasma. The non resonant event is a local event driven in the middle plasma and its destabilizing effect in the inner plasma is small. In summary, the non resonant event is a system relaxation in the soft MHD limit because it cant be the driver of a plasma collapse, but in the case of the resonant event it is a relaxation in the hard MHD limit if it is strong enough to trigger a collapse.

To study the role of the resonant events in the transition between the soft-hard MHD regimes for a plasma with different stability properties, we increase the value of the Lundquist number in the next simulations.

\subsection{Simulation with $S = 5 \times 10^5$}

Again, we show a chain of a non resonant (I) and a resonant (II) $1/3$ sawtooth like event. The main deformation of the pressure profile during the non resonant event is a flattening in the middle plasma driven by the mode $1/2$, $t = 0.2802$ and $0.2804$ s (Fig. 8 A). At $t = 0.2806$ s a new profile flattening appears around $\rho = 0.3$ driven by the modes $2/5$ and $3/7$. After the onset of the resonant event the pressure profile drops very close to the magnetic axis by the destabilizing effect of the mode $1/3$, $t = 0.2808$ s, and it is followed by a large profile drop in all the inner plasma region, $t = 0.2810$ s. The iota profile in the non resonant event is deformed in the inner plasma region, reaching a local minimum around $\rho = 0.14$ but it doesn't drop below $\rlap{-} \iota = 1/3$. During the resonant event the profile deformation is stronger, specially close to the magnetic axis where it drops below $\rlap{-} \iota = 1/3$ at $t = 0.2808$ s and the destabilizing effect of the other unstable modes in the inner plasma is large.

\begin{figure}[h]
\centering
\includegraphics[width=0.47\textwidth]{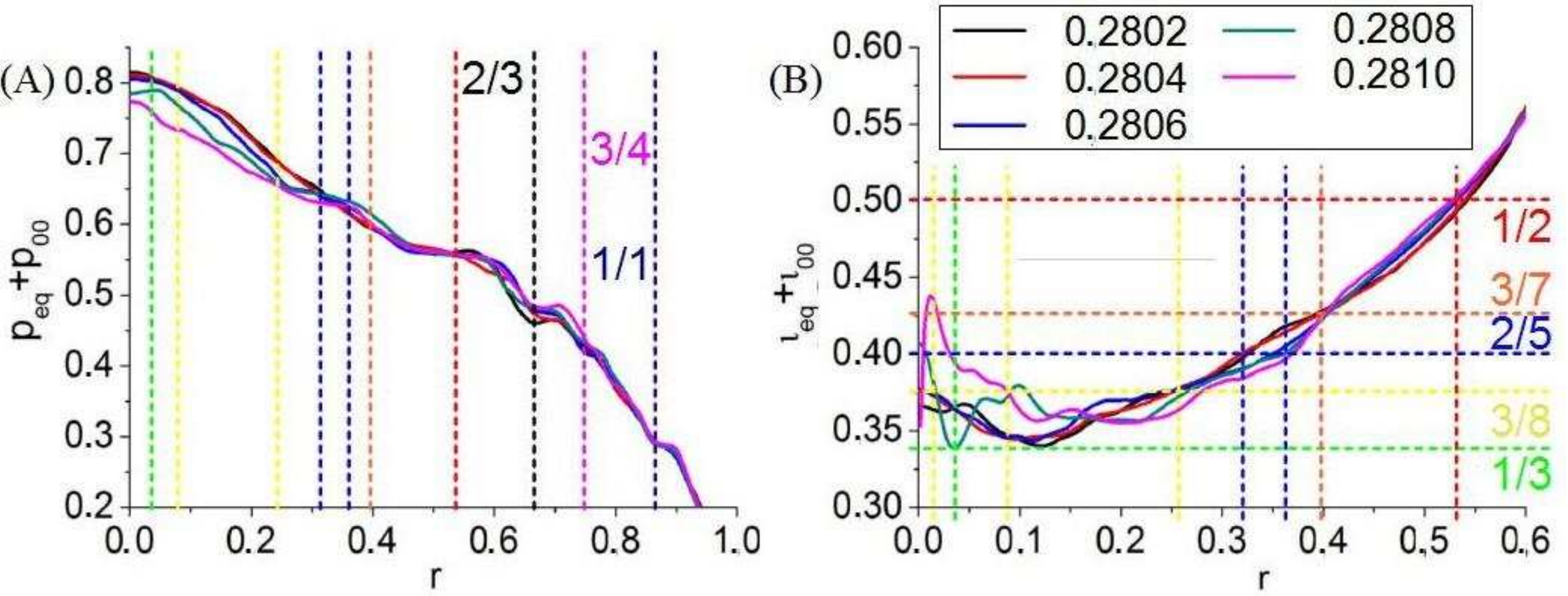}
\caption{Pressure profile (A) and the rotational transformation (B) in the simulation with $S = 5 \times 10^5$.} \label{FIG:8}
\end{figure} 

The flux surfaces during the non resonant event are deformed in the middle plasma region, $t = 0.2802$ and $0.2804$ s (Fig. 11A). At $t = 0.2806$ and $0.2808$ s a large perturbation appears in the inner plasma that is enhanced during the resonant event, leading to a large torn of the flux surfaces in the inner plasma and a plasma leak to the outer torus. At $t = 0.2810$ s the flux surfaces of the inner and middle plasma are almost disconnected. There is a stochastic region between the middle and the inner plasma in the non resonant case, but it doesn't reach the magnetic axis (Fig. 11B). During the resonant case there is a large stochastic region covering all the inner plasma that reaches the magnetic axis, but it is not linked with the stochastic region in the middle plasma. The magnetic islands between the middle and inner plasmas are overlapped during the non resonant event but the magnetic surfaces close to the magnetic axis are not distorted (Fig. 11C). In the resonant case the $1/3$ islands appear near the magnetic axis and the magnetic islands in the inner plasma are strongly overlapped.

\begin{figure}[h]
\centering
\includegraphics[width=0.48\textwidth]{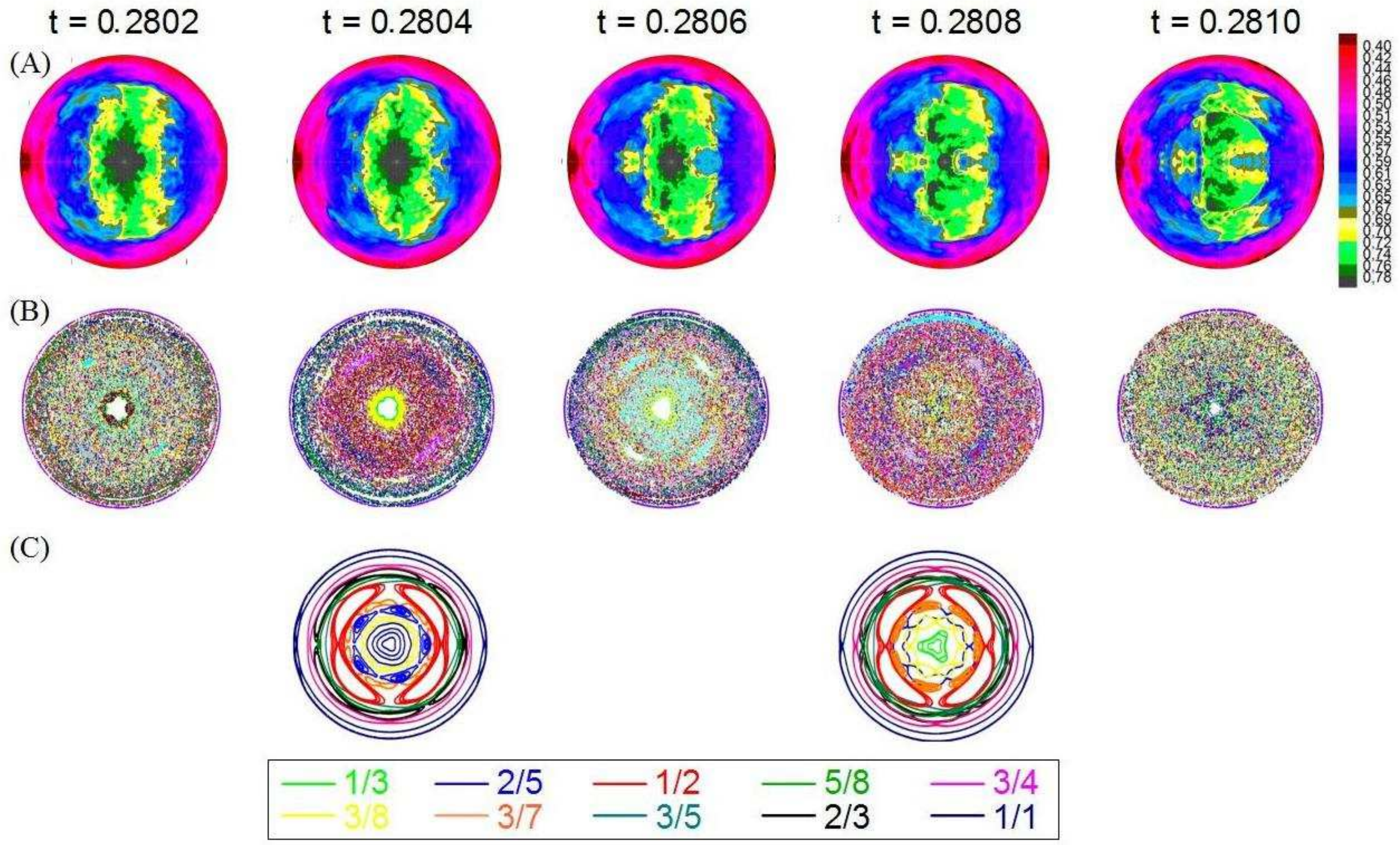}
\caption{Contour plot of the perturbed pressure (A), magnetic islands (B) and stochastic region (C) in the simulation with $S = 5 \times 10^5$.} \label{FIG:9}
\end{figure} 

These results point out that there is a soft-hard transition of the MHD regime after the onset of the $1/3$ resonant sawtooth like event. When the $1/3$ magnetic islands appear the stochastic region reaches the magnetic axis, covering all the inner plasma. The pressure profile drops and the iota profile is strongly distorted in the inner region as well as the flux surfaces. The inner plasma collapses leading to a large plasma and energy leak.

\subsection{Simulation with $S = 10^6$}

This simulation shows a chain of one non resonant event (I) and two resonant events (II and III). The pressure profile evolution shows a flattening in the middle plasma driven by the mode $1/2$ and another two flattening in the outer plasma driven by the modes $2/3$, $3/4$ and $1/1$ (Fig. 14A). There isn't a profile flattening in the inner plasma, only a small drop near the magnetic axis driven by the mode $1/3$. The iota profile is deformed close to the magnetic axis but it is weaker than in the previous simulation (Fig. 14B). The iota profile drops below $1/3$ around $\rho = 0.1$.

\begin{figure}[h]
\centering
\includegraphics[width=0.47\textwidth]{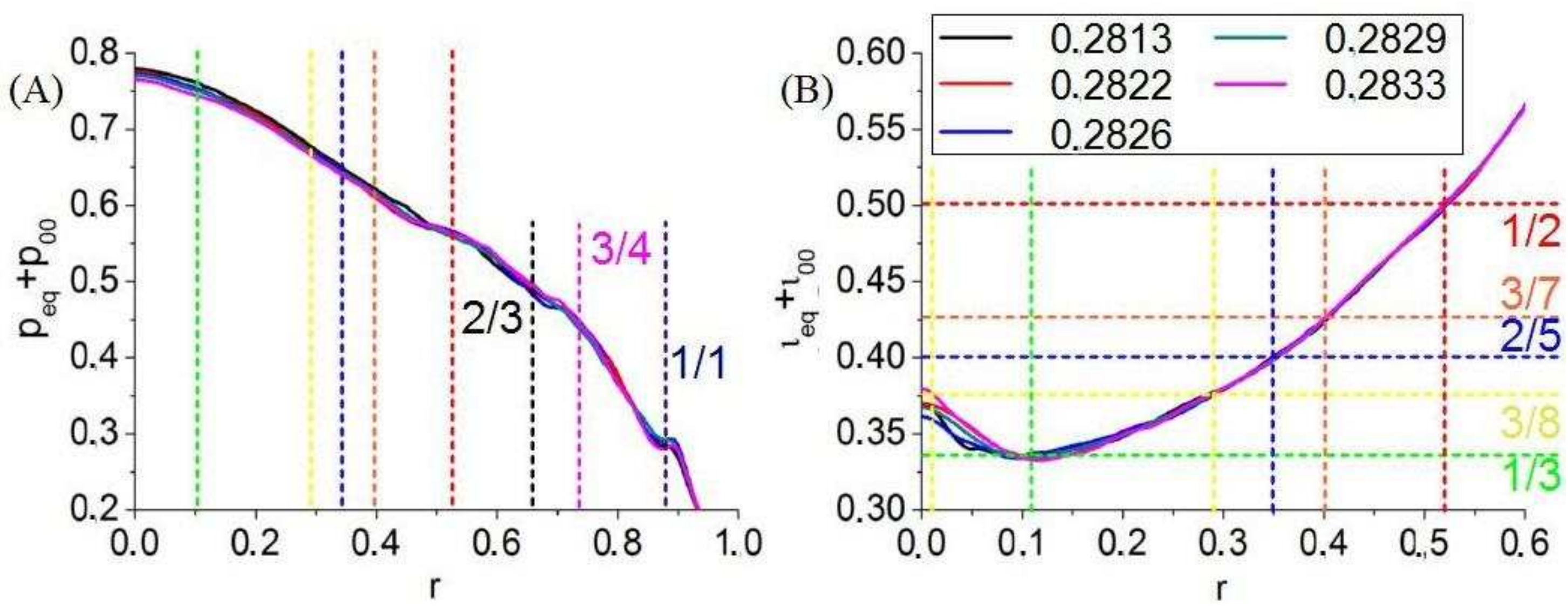}
\caption{Pressure profile (A) and the rotational transformation (B) in the simulation with $S = 10^6$.} \label{FIG:10}
\end{figure} 

The flux surface between the middle and outer plasma are deformed but not torn during the three events. They are slightly distorted in the second and third events in the inner plasma, $t = 0.2826$ and $0.2833$ s (Fig. 15A). During the first event there is a stochastic region between the middle and outer plasma, but in the second and third event there is another stochastic region in the inner plasma, not linked with the stochastic region in the middle plasma (Fig. 15B). The magnetic surfaces in the inner plasma are only slightly perturbed because the destabilizing effect of the modes and the width of the magnetic islands is small (Fig. 15C), like the weak effect of the $1/3$ islands near the magnetic axis.

 \begin{figure}[h]
\centering
\includegraphics[width=0.48\textwidth]{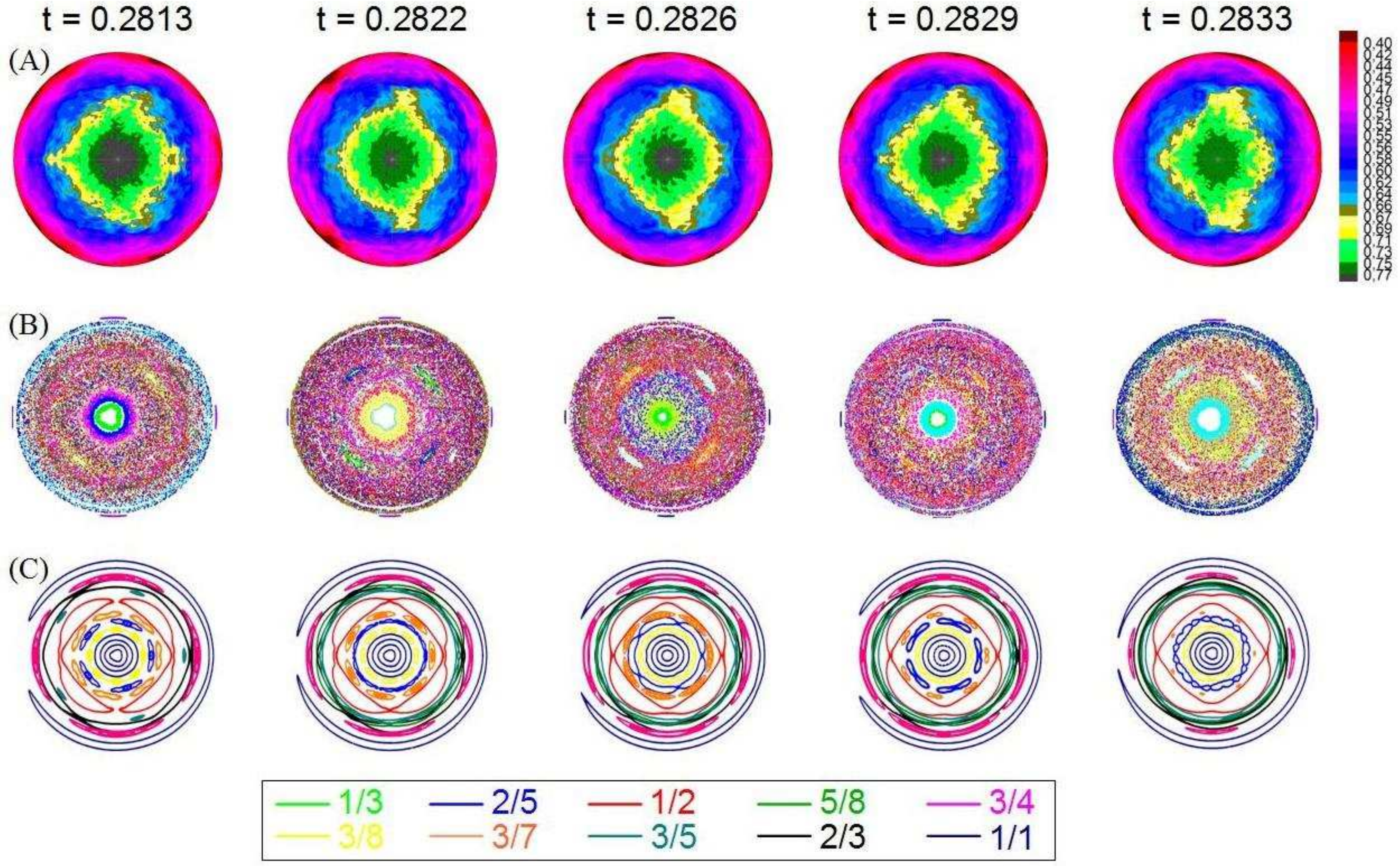}
\caption{Contour plot of the perturbed pressure (A), magnetic islands (B) and stochastic region (C) in the simulation with $S = 10^6$.} \label{FIG:11}
\end{figure} 

This chain of events are system relaxations in the soft MHD regime. The overlapping between the magnetic islands in the inner plasma is weaker than in the previous simulation. The main difference between the resonant events in this simulation and the simulation with $S = 5 \times 10^5$ is the destabilizing effect of the $1/3$ mode near the magnetic axis. The width of the $1/3$ islands is small and the stochastic region in the inner plasma doesn't reach the magnetic axis, therefore the resonant events don't drive a collapse in the inner plasma and the system remains in the soft MHD limit.

\subsection{Comparative study}

To study the plasma stability properties in each MHD regime and why the transition is triggered, we compare the evolution of the energy loss, the normalized ME, the single mode ME, the magnetic turbulence and the pressure gradient of the simulations with $S = 5 \times 10^5$ and $S = 10^6$. 

The energy loss in the $S = 5 \times 10^5$ simulation is two times higher in the resonant event than in the non resonant event (Fig. 12A). The energy loss in the three events of the $S = 10^6$ case is similar to the non resonant case of the simulation with $S = 5 \times 10^5$ (Fig. 12B). This means that the resonant relaxation of the $S = 5 \times 10^5$ simulation is the strongest event and it drives the largest drop in the system capacity to confine the plasma.

\begin{figure}[h]
\centering
\includegraphics[width=0.35\textwidth]{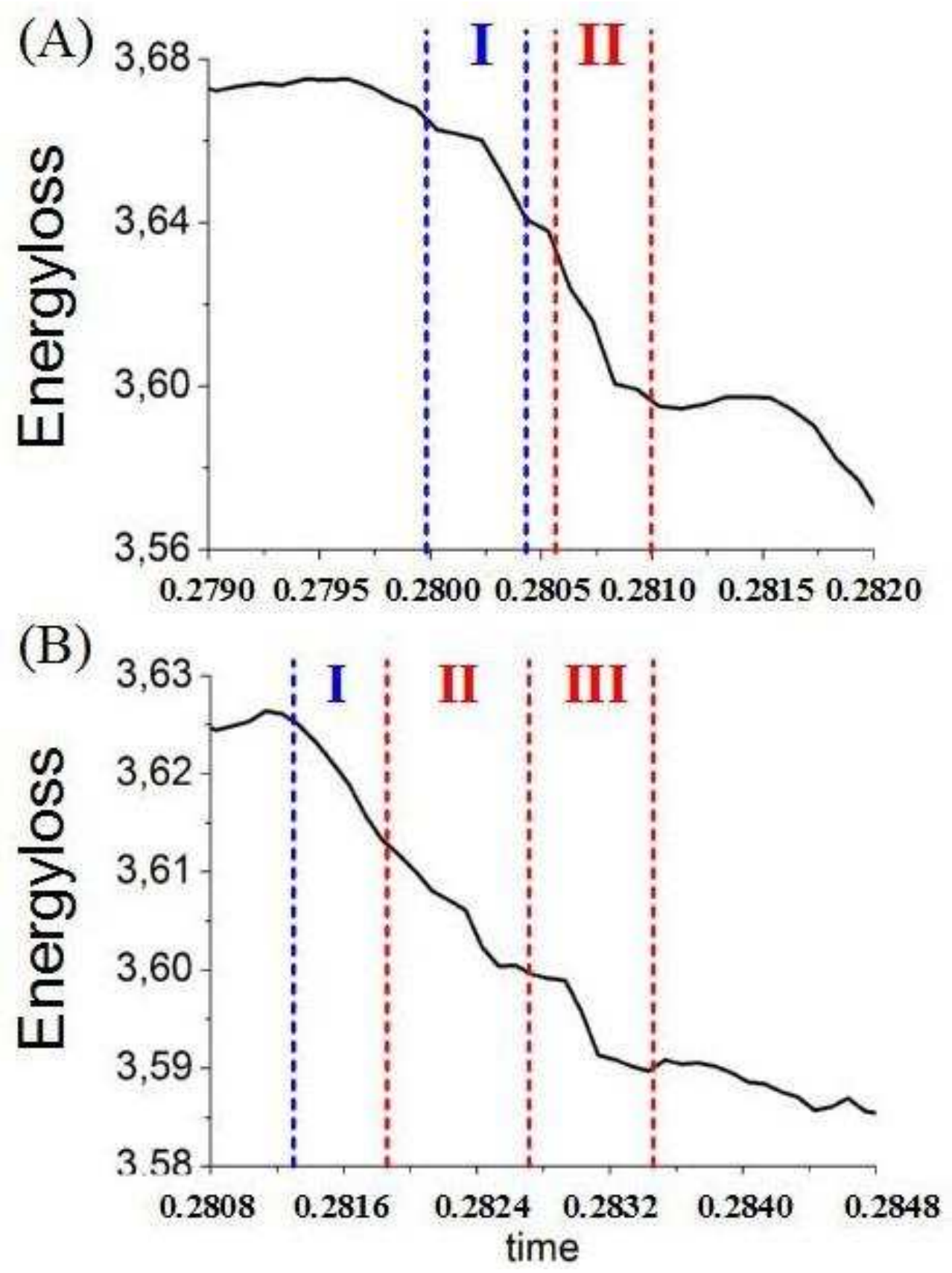}
\caption{Evolution of the system energy loss for the $S = 5 \times 10^5$ (A) and $S = 10^6$ (B) simulations.} \label{FIG:12}
\end{figure} 

The maximum of the system ME in the $S = 5 \times 10^5$ simulation is reached during the resonant event and it is two times larger than the local maximum in the non resonant event (Fig. 13A). There are two local maximum in the $S = 10^6$ simulation during the second and third resonant events, but these maximums are small compared with the $S = 5 \times 10^5$ case (Fig. 13B). The ME evolution indicates too that the resonant event in the $S = 5 \times 10^5$ simulation is the strongest relaxation. The effect of the $S = 10^6$ resonant events in the ME evolution is similar to the non resonant event in the $S = 5 \times 10^5$. 

\begin{figure}[h]
\centering
\includegraphics[width=0.35\textwidth]{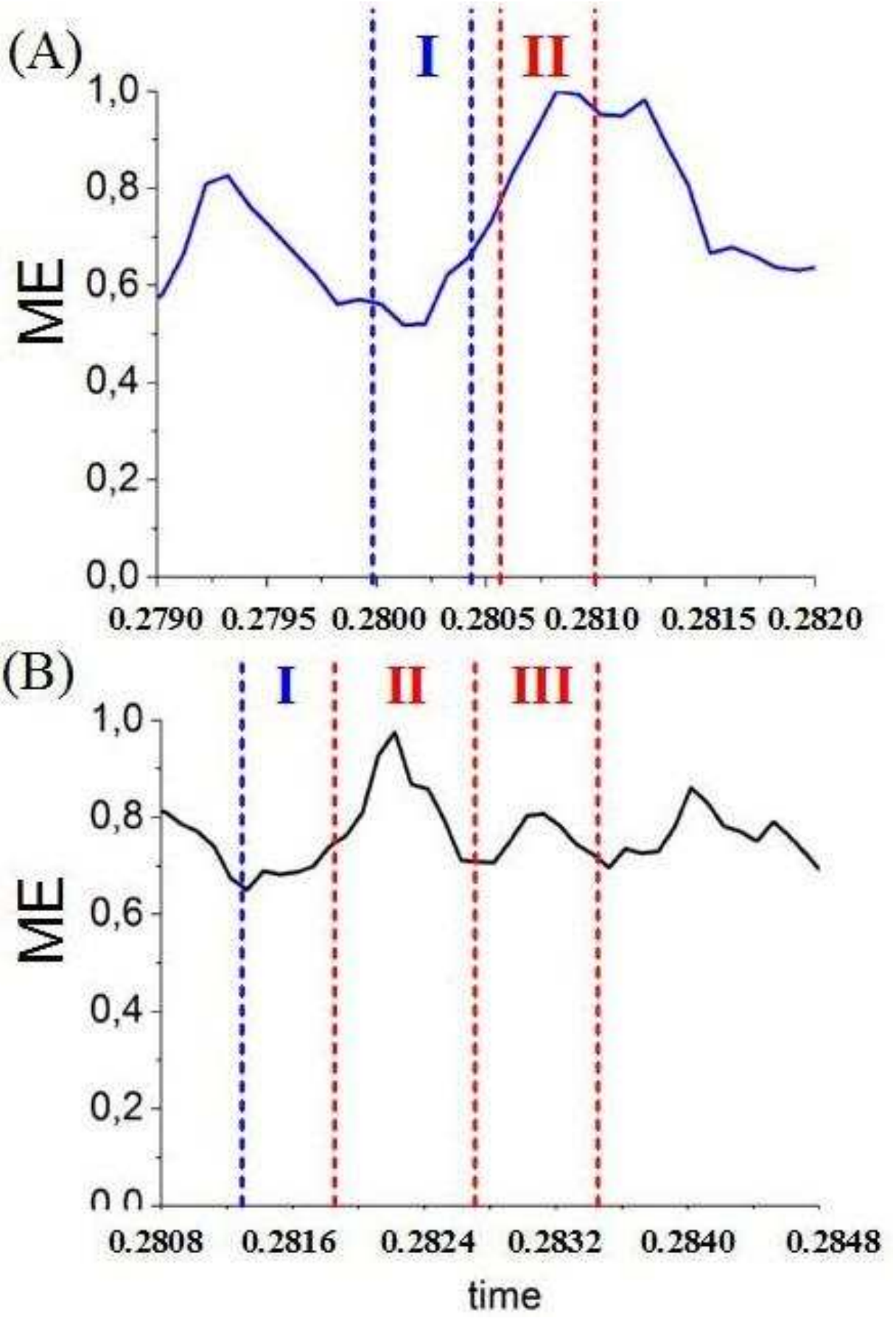}
\caption{Evolution of the normalized ME for the $S = 5 \times 10^5$ (A) and $S = 10^6$ (B) simulations.} \label{FIG:13}
\end{figure} 

The ME evolution of the $1/3$ mode in the $S = 5 \times 10^5$ simulation is similar to the original case with $S = 10^5$, it grows during the non resonant event and it reaches a local maximum in the resonant event (Fig. 14A). There is a strong correlation in the evolution of the inner plasma modes ME, specially in the resonant case. For the $S = 10^6$ simulation the local maximum of the energy for the modes in the inner plasma is small compared with the energy of the modes in the middle and outer plasma region (Fig. 14B). These results point out that the inner plasma region is more unstable in the $S = 5 \times 10^5$ simulation and the resonant relaxation affects all the inner plasma region.

\begin{figure}[h]
\centering
\includegraphics[width=0.4\textwidth]{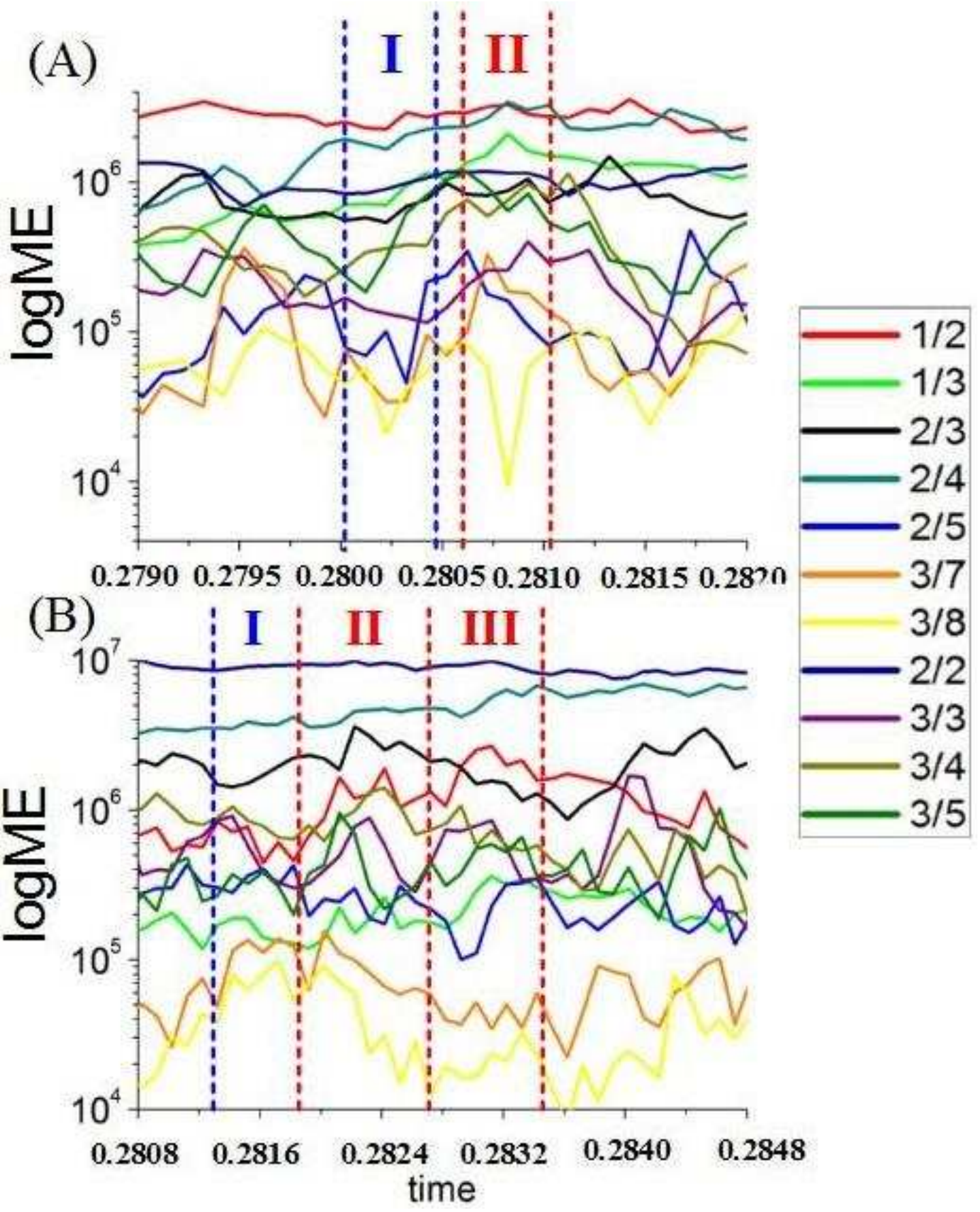}
\caption{Evolution of the single mode ME for the $S = 5 \times 10^5$ (A) and $S = 10^6$ (B) simulations.} \label{FIG:14}
\end{figure} 

The magnetic turbulence in the $S = 5 \times 10^5$ simulation begins to grow in the non resonant event and it reaches a local maximum during the resonant event (Fig. 15A). The averaged value of the magnetic turbulence in the $S = 10^6$ simulation is higher but the local maximum reached in the third event is smaller, so the gradient of the magnetic turbulence is lesser (Fig. 15B). This means that there is a correlation between the magnetic turbulence and the triggering of the soft-hard MHD transition. 

\begin{figure}[h]
\centering
\includegraphics[width=0.35\textwidth]{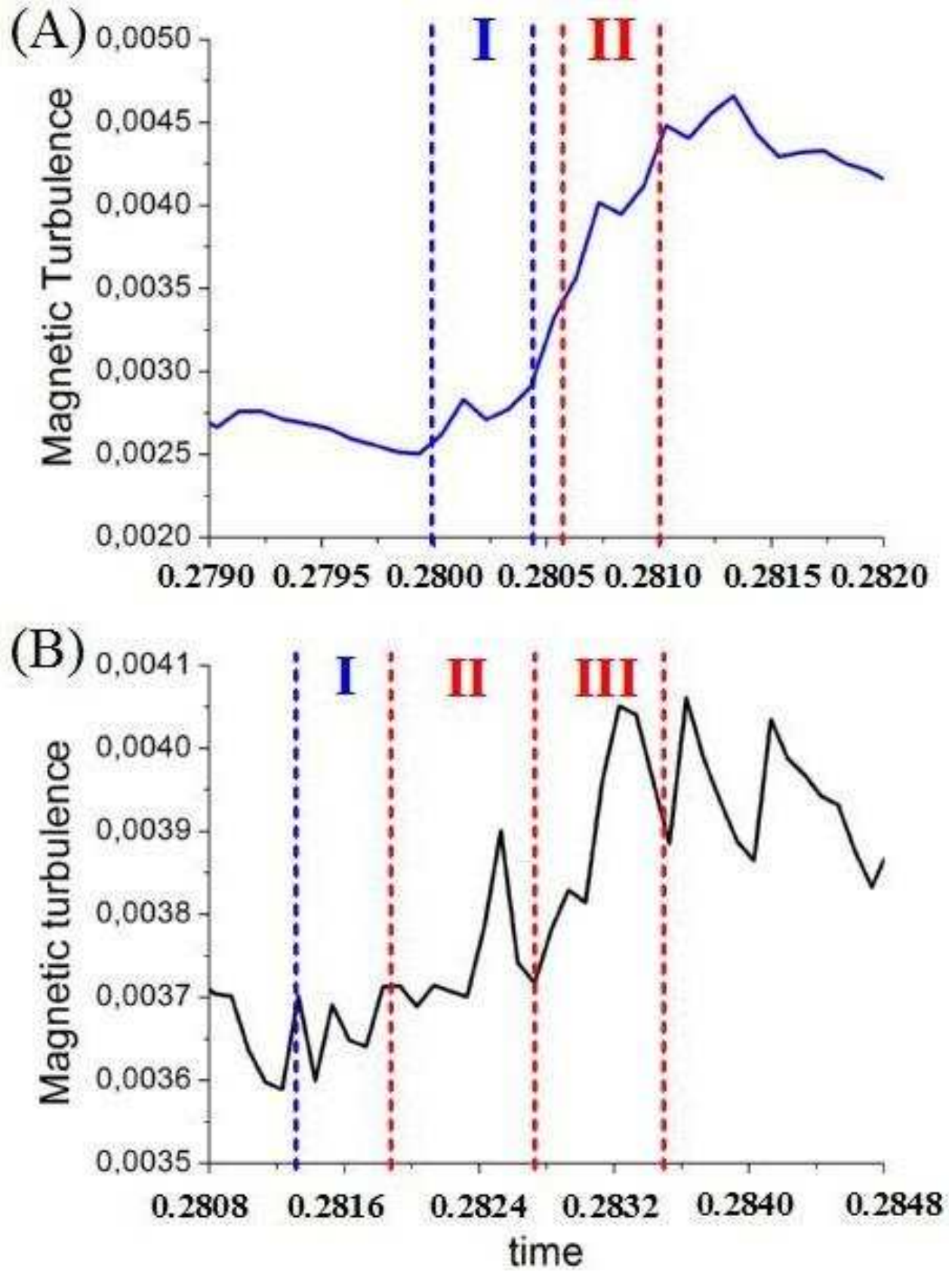}
\caption{Evolution of the magnetic turbulence for the $S = 5 \times 10^5$ (A) and $S = 10^6$ (B) simulations.} \label{FIG:15}
\end{figure} 

During the $S = 5 \times 10^5$ simulation there are small peaks of the pressure gradient profile in the inner plasma when the non resonant event is driven, as well as large oscillations just before the onset and during the resonant event (Fig 16A). In both relaxations the pressure gradient in the middle and outer plasma keeps almost constant. Along the first event of the $S = 10^6$ simulation, there is a peak of the pressure gradient profile in the middle plasma at $t = 0.2811$ s, and in the inner plasma at $t = 0.2816$ s, therefore the perturbation origin is the middle plasma and it expands to the inner plasma (Fig. 16B). During the second and third event there is a perturbation between $\rho = 0.1$ and $0.3$ at $t = 0.2822$ and $0.2832$ s and it reaches the nearby of the middle plasma region at $t = 0.2824$ and $0.2834$ s, hence the instability origin is the inner plasma. The perturbation during the second event is stronger in the middle plasma but in the third event it is stronger in the inner plasma. These results show that the pressure gradient is higher in the $S = 5 \times 10^5$ case, specially in the inner plasma around $\rho = 0.1$ during the resonant event.

\begin{figure}[h]
\centering
\includegraphics[width=0.3\textwidth]{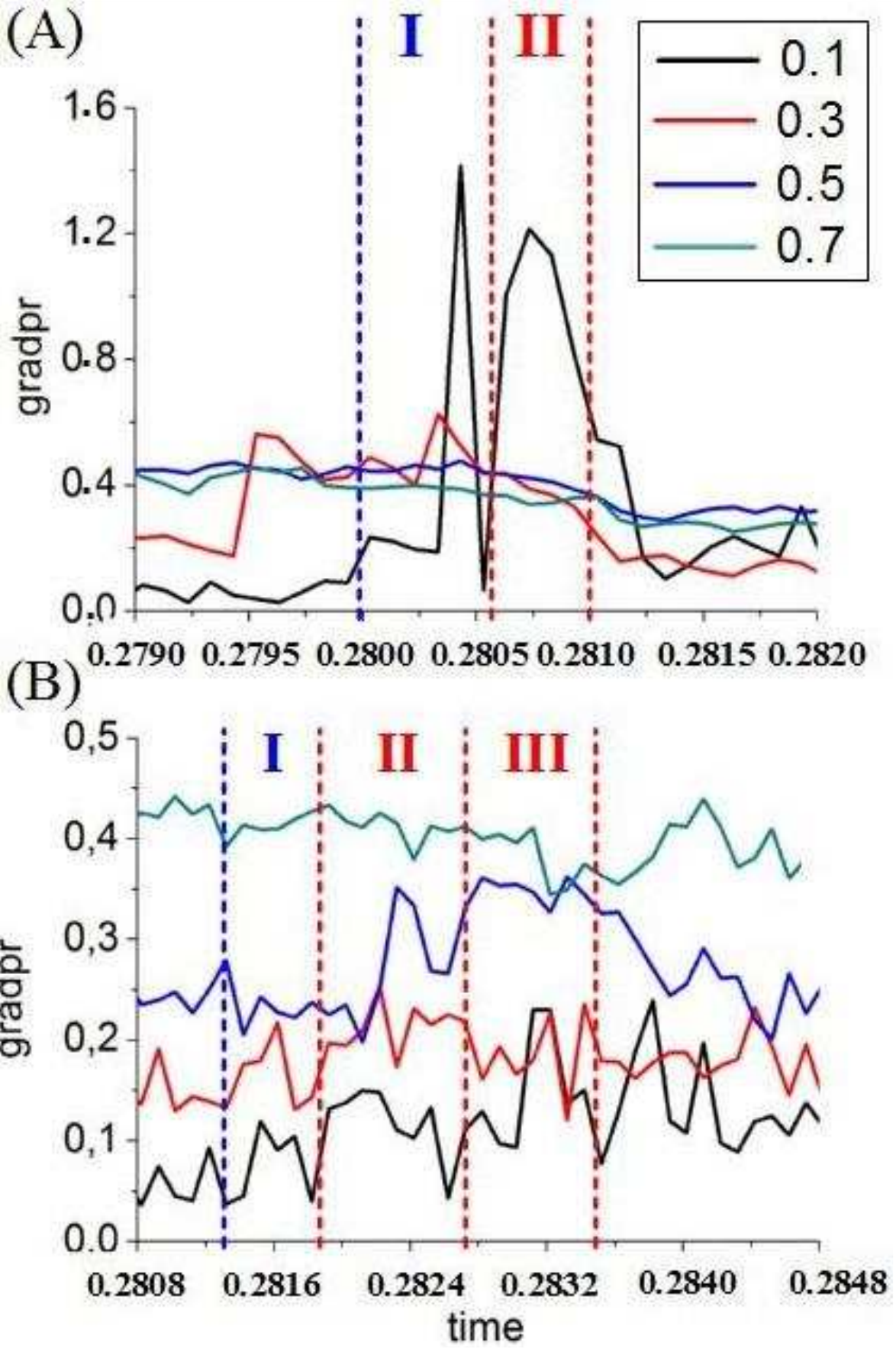}
\caption{Evolution of the pressure gradient for the $S = 5 \times 10^5$ (A) and $S = 10^6$ (B) simulations.} \label{FIG:16}
\end{figure} 

In summary, during the simulation with $S = 5 \times 10^5$ the pressure gradient and the magnetic turbulence are high enough to drive the transition to the hard MHD regime. The collapse of the inner plasma region is triggered after the onset of a strong $1/3$ resonant sawtooth like event. The plasma of the $S = 10^6$ simulation remains in the soft MHD regime because the pressure gradient and the magnetic turbulence are below the critical value to drive the soft-hard MHD limit transition. The resonant events driven during the simulation are weak and the collapse is not triggered.

\section{Conclusions and discussion \label{sec:conclusions}}

There is a transition from the soft to the hard MHD regime in the simulation with $S = 5 \times 10^5$. The inner plasma collapses if the overlapping of the magnetic islands $1/3$, $3/8$, $2/5$ and $3/7$ is strong. A stochastic region covers the inner plasma including the magnetic axis, but it is not linked with the stochastic region in the middle plasma. The collapse is driven after the onset of a strong resonant $1/3$ sawtooth like event, while there is a maximum of the magnetic turbulence and the pressure gradients in the inner plasma. The iota profile is strongly deformed and the destabilizing effect of the modes in the inner plasma reach the region close to the magnetic axis. The pressure profile drops near the magnetic axis and the flux surfaces are torn in the inner and middle plasma.

There isn't a transition between the soft-hard MHD regimes in the simulation with $S = 10^6$. Two resonant $1/3$ events are driven but they are weaker than in the simulation with $S = 5 \times 10^5$. The width of the $1/3$ islands is small and they are not overlapped with the other islands in the inner plasma, therefore the stochastic region doesn't reach the magnetic axis. The magnetic turbulence and pressure gradient in the inner plasma are not large enough to drive a strong magnetic island overlapping and no collapse is observed. The iota profile drops below $\rlap{-} \iota = 1/3$ but  the destabilizing effect of the modes in the inner plasma region is small near the magnetic axis. The pressure profile doesn't show a large drop or flattening in the inner plasma and the flux surfaces are only slightly deformed and never torn.

A collapse in the inner plasma can be driven after the onset of a strong $1/3$ resonant event if the magnetic islands overlapping is large, and a large stochastic regions covers all the inner plasma including the magnetic axis. The width of the magnetic islands is critical, so a plasmas with a large magnetic turbulence and pressure gradient in the inner plasma can collapse easily. This is the case of the simulation with $S = 5 \times 10^5$, there is a transition from the soft to the hard MHD regime, but not in the $S = 10^6$ simulation that remains in the soft MHD regime.

The onset of the collapse can be avoided if the $1/3$ resonant sawtooth like events are not driven. No collapse is observed in the inner plasma if the system relaxation is a non resonant $1/3$ event. If a resonant events is triggered but the width of the magnetic islands in the inner plasma is not large enough to be strongly overlapped, the system remains in the soft MHD limit. This is the case if the magnetic turbulence and the pressure gradient in the inner plasma is below the critical value to drive the soft-hard MHD transition. For LHD operations with high density and low resistivity in the inner plasma, the magnetic turbulence is small and the system remains easily in the soft MHD limit.

The distortion of the iota profile in the inner plasma is another important issue in the soft-hard transition. If the deformation of the iota profile is large, the modes in the inner plasma can destabilize the region close to the magnetic axis, distorting the magnetic surfaces and enhancing the islands overlapping, therefore the system can reach the hard MHD limit easily. In LHD operations with a non negligible current in the inner plasma, normally induced by a strong NBI heating, or in operations with net toroidal currents, the iota profile can be strongly distorted. In these LHD operations the transition to the hard regime are easily attain as well as the onset of a collapse in the inner plasma.

\begin{acknowledgments}
This study is supported by the SHOCK european project and the group LESIA of the Observatorie Paris-Meudon. The authors are very grateful to L. Garcia for letting us use the FAR3D code and his collaboration in the developing of the present manuscript diagnostics.
\end{acknowledgments}

\end{document}